\documentclass [12pt]{article}
\usepackage {amssymb}
\usepackage {amsmath}
\usepackage {graphicx}

\begin{document}

\begin{flushleft}
{\footnotesize{arXiv.org e-print archive cond-mat/0108417}}
\end{flushleft}

\begin{center}
\section*{Collective dynamics of hydrogen atoms in the
KIO$_3 \cdot$HIO$_3$ crystal dictated by a substructure of the
hydrogen atoms' matter waves}
\end{center}

\author{{\bf Volodymyr Krasnoholovets\footnote{e-mail:
krasnoh@iop.kiev.ua}} \\
 {} \\
Institute of Physics, National Academy of Sciences \\
 Prospect Nauky 46,  UA-03028 Ky\"{\i}v, Ukraine \\
 http://inerton.cjb.net}
\date{23 January 2003}

\bigskip

\begin{abstract}
The behavior of the subsystem of hydrogen atoms of the KIO$_3
\cdot$HIO$_3$ crystal, whose IR absorption spectra exhibit
equidistant submaxima in the vicinity of the maxima in the
frequency range of stretching and bending vibrations of OH bonds
is studied in the present work. It is shown that hydrogen atoms
co-operate in peculiar clusters in which, however, the hydrogen
atoms do not move from their equilibrium positions, but become to
vibrate synchronously. The interaction between the hydrogen atoms
is associated with the overlapping their matter waves, or more
exactly, the waves' substructure, a very light quasi-particles
called {\it inertons}, which play the basic role in submicroscopic
quantum mechanics that has recently been constructed by the
author. The exchange by inertons results in the oscillation of
hydrogen atoms in clusters, which emerges in the mentioned
spectra. The number of atoms, which compose the cluster, is
calculated.

\end{abstract}

\textbf{PACS:} 03.75.-b Matter waves; \ 34.10+x General theories
and models of atomic and molecular collisions and interactions
(including statistical theories, transition state, stochastic and
trajectory models, etc.); \  14.80.-j Other particles (including
hypothetical)

\medskip

\textbf{Key words:} statistical mechanics, clusters, hydrogen
structures, interparticle interaction, matter waves, inertons, IR
spectra

\medskip

\section*{1. Statement of the problem}

\hspace*{\parindent} This work presents a manifold research, which
embraces the experimental results by Puchkovska and collaborators
[1]  on a very peculiar behavior of hydrogen atoms in the
$\delta$-KIO$_3 \cdot$HIO$_3$ crystal, a rather new concept of the
clustering, which appears at the statistical mechanical
description of interacting particles, first proposed by Belotsky
and Lev [2] (see also Refs. [3-5]) and submicroscopic quantum
mechanics developed in the real space, which has recently been
constructed by the author [6-13].

Below the major experimental results obtained [1,14-16] on the
delta modification of the KIO$_3 \cdot$HIO$_3$ crystal is
described. In particular, the FT-IR absorption spectra of powdered
samples in the frequency range of stretching and bending
vibrations of OH bonds are analyzed. Since the spectroscopic
studies reveal a modulation of the spectral bands, this effect is
associated with collective vibrations of hydrogen atoms. Such a
behavior is possible only in the case when the hydrogen atoms
represent a dynamic system of a sort. To study this possibility,
we should first have the rigorous evidence that the behavior of
hydrogen atoms, or protons are distinguished from the backbone
atoms. Note that this the subject of study that Fillaux and
collaborators have been conducting for decade [17-22]: Using the
incoherent inelastic neutron scattering technique they have
investigated vibrational dynamics for protons in various solids
and revealed that proton dynamics is almost totally decoupled from
surrounding heavy atoms. Besides, Fillaux noted that the proton
subsystem demonstrates a collective dynamics (see also Trommsdorff
et al. [23,24], who disclosed the coherent proton tunneling and
cooperative proton tunneling and transfer of four protons in
hydrogen bonds of benzoic acids crystals).

The statistical mechanical approach proposed in Refs. [2-5] makes
allowance for spatial nonhomogeneous states of interacting
particles in the system studied. However, in the case when the
inverse operator of the interaction energy cannot be determined
one should search for the other method, which, nevertheless, will
make it possible to take into account a plausible nonhomogeneous
particle distribution. In papers [2-5] systems of interacting
particles have been treated from the same standpoint [1], however,
the number of variables describing the systems in question has
been reduced and a new canonical variable, which characterized the
nascent nonhomogeneous state (i.e. cluster), automatically aroused
as a logical consequence of the  behavior of particles. A detailed
analysis of the appearance of nonhomogeneous states in systems of
interacting particles submitted to quantum statistics (Fermi's and
Bose') has been performed in Ref. [5]. In the present work we will
consider a possible clustering of hydrogen atoms in the KIO$_3
\cdot$HIO$_3$ crystal in the framework of Boltzmann statistics
based on ideas developed in Refs. [2-5]. If we are able to form a
cluster of hydrogen atoms, this automatically will mean that the
hydrogen atoms fall within proper vibrations, which should be
distinguished from the spectra of heavy atoms. Thus, if clusters
of hydrogen atoms exist, each of them  will represent a typical
dynamic system and therefore vibrations of hydrogen atoms being
superimposed on the carrying absorption OH bands will induce a
modulation of the OH bands, which has been reveled in the
experiment [1].

What kind of a problem can one meet at such a consideration? The
said approach can bring about clusters only in the case when
repulsive and attractive components of the pair potential feature
very different dependencies on distance from the particle.
However, we can not expect that the electromagnetic interaction
between hydrogen atoms in the crystal can exhibit a great
difference between repulsive and attractive components of the pair
potential (for example, for the Van der Waals interaction the
behavior of the components, respectively $1/r^{12}$ and $1/r^6$,
are not fundamentally different).

It has been shown in the author works devoted to the foundations
of quantum mechanics [6-13] that the matter waves of particles are
characterized by a certain substructure, namely, that the matter
waves come from the interaction of a moving particle with a space
substrate, i.e. real space. Due to the interaction, an ensemble of
sub quasi-particles  should appear surrounding the particle (these
quasi-particles  were called "inertons" [6], as they represent
inert properties of canonical particles). A cloud of inertons
spreads around the particle in the range
\begin{equation}
\label{1} \Lambda = \lambda {\kern 1pt} {\kern 1pt} c/v
\end{equation}
where $\lambda $ is the particle's de Broglie wavelength, $v$ and
\textit{c} are velocities of the particle and light, respectively.
Thus the wave $\psi$-function becomes determined in the range
around the particle covered by the amplitude of inerton cloud (1).
Submicroscopic mechanics developed in the real space [6-13] makes
it possible to overcome all conceptual difficulties inherent in
orthodox statistical quantum mechanics (developed in an abstract
space) and at the same time submicroscopic mechanics has easily
fitted into Schr\"odinger's [6,7] and Dirac's [9] formalisms
accounting for the inner reason of allegedly different
prerequisites for the two approaches [9,12]. A great number of
experimental manifestations of clouds of inertons surrounding
electrons was demonstrated in Ref. [10]. Our own prediction and
experimental verification of the existence of inertons were stated
in paper [8].

In the present work we shall use one consequence, which directly
follows from the theory [6-13] and which has already been employed
in papers [5,8]. Clouds of inertons expended around the same
particles (for instance, hydrogen atoms) are exemplified by the
same characteristics, namely: amplitude (1), mass, momentum, and
energy. This means that a system of such clouds should interact
much as elastic balls and, therefore, the potential energy of the
interacting clouds may be written as $\gamma{\kern 1pt} r^2/2$
where $r$ is a deviation of an atom from its equilibrium position
in the lattice and $\gamma$ is an elasticity, or force, constant
of the lattice's inerton field.

Keeping in mind the aforesaid, let us now turn to a close
examination of the behavior of hydrogen atoms in the KIO$_3
\cdot$HIO$_3$ crystal.

\section*{2. Peculiarities experimentally revealed
in hydrogen atoms dynamics}

\hspace*{\parindent}  The crystal structure, proton disorder and
low temperature phase transition of the $\delta$-KIO$_3
\cdot$HIO$_3$ crystal have been studied by means of X-ray
diffraction, dielectric, calorimetric and FT-IR and FT-Raman
techniques by Engelen et al. [14]. It has been found that
[I$_{3}$O$_{9}$H$_{3/2}$]$^{-3/2}$ ions are linked by means of
hydrogen bonds to the oxygen atom at the middle of two other
[I$_{3}$O$_{9}$H$_{3/2}$]$^{-3/2}$ anions. Moreover,
hydrogen-bonded plane grids parallel (100) are formed with
hydrogen bonded chains in [011]. K$^{+}$ ion is located between
the plane grids and completes the crystal structure of
$\delta$-KIO$_3 \cdot$HIO$_3$. Basically, in the crystal studied
hydrogen bonds do not form an entire uninterrupted network, but
rather represent local islands and each of them features a compact
ordered system of hydrogen bonds.

Further experimental studies conducted by Puchkovska and
collaborators [14,1] have been aimed at the detailed measurements
of the FT-IR absorption spectra of the powdered $\delta$-KIO$_3
\cdot$HIO$_3$ crystal in the frequency range of stretching and
bending vibrations of OH bonds for the temperature range from 13
to 293 K. The spectra recorded in high-frequency region are shown
in Figs. 1 and 2. Hydrogen bonds in the crystal under
consideration were ascribed to weak and medium hydrogen bonds. For
such hydrogen bonds Fermi resonance does not practically excite
$\nu$(OH) spectral band shape [25]. However, the presence of
K$_{3/2}$[I$_{3}$O$_{9}$H$_{3/2}$] moieties is the major
structural feature of the $\delta$-KIO$_{3}\cdot$HIO$_{3}$, which
are able to induce a similar distribution observed in the
$\nu$(OH) band for $\alpha$-HIO$_{3}$ crystal [16].

\begin{figure}
\begin{center}
\includegraphics[scale=0.48]{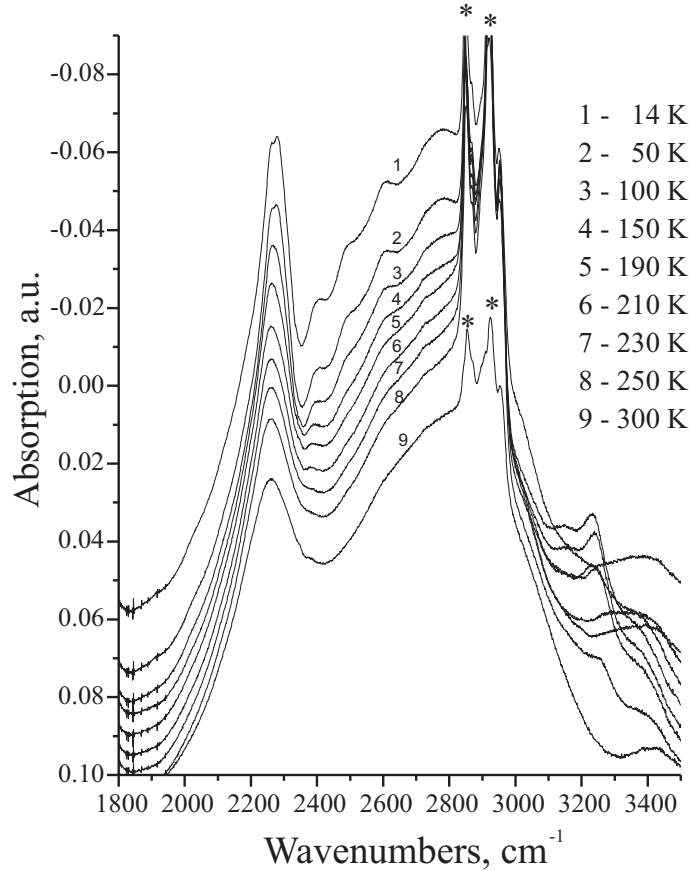}
\caption{Temperature dependent IR spectra of polycrystalline
$\delta$-KIO$_{3}\cdot$HIO$_{3}$ (suspension in Nujol) in the
region of stretching $\nu$(OH) mode. Asterics denote the Nujol
absorption bands. (With  permission from Ref. [1].)} \label{Figure
1}
\end{center}
\end{figure}

\begin{figure}
\begin{center}
\includegraphics[scale=0.63]{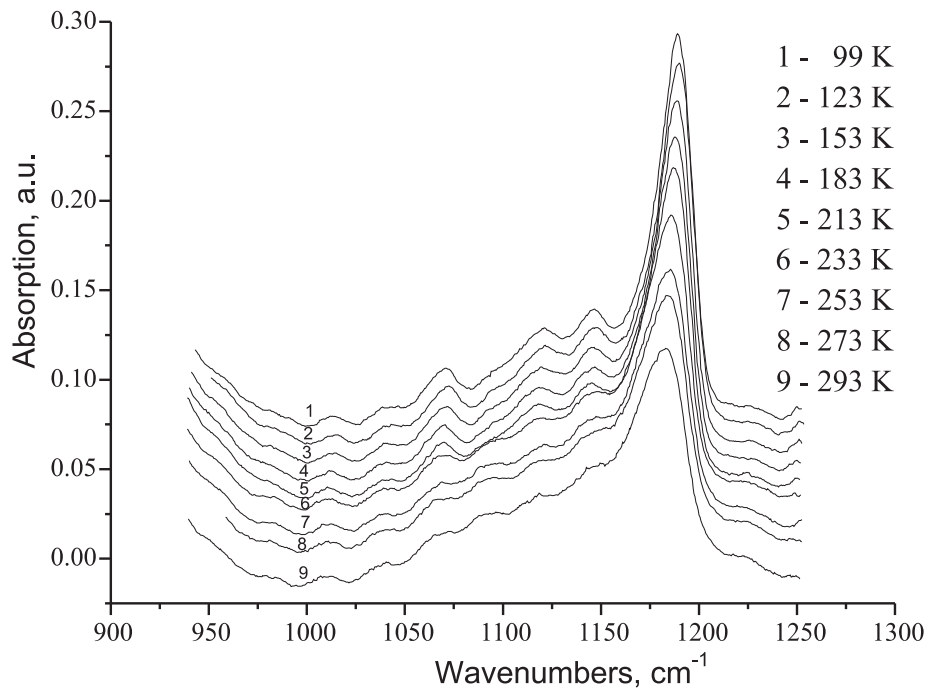}
\caption{Temperature dependent IR spectra of polycrystalline
$\delta$-KIO$_{3}\cdot$HIO$_{3}$ in NaCl pellet in the region of
the in-plane bending mode $\delta$(OH).(With  permission from Ref.
[1].)} \label{Figure 2}
\end{center}
\end{figure}

A very strong broad band centered at 2900 cm$^{-1}$ and less
intensive band at 2320 cm$^{-1}$ have been assigned [14] to
$\nu$(OH) stretching vibration and $2\delta$(OH) overtone,
respectively. In the 1000 to 1300 cm$^{-1}$ region there is a band
of the medium intensity with peak position 1185 cm$^{-1}$ at the
room temperature; the frequencies belong to the in-plane bending
modes $\delta$(OH) of I -- O -- H$\cdot \cdot \cdot $O -- I
fragments. The temperature dependencies of the peak position and
half-width of the $\delta$(OH) band were measured in work [16] and
showed sharp changes in their slopes near 220 K. This was
associated with the phase transition in the proton sublattice of
order-disorder type.

The most interesting result reported in Refs. [14,1] is the
disclosure of a pronounced "modulation" in both low-frequency
slopes and high-frequency ones of the $\nu$(OH) and $\delta$(OH)
bands especially at temperature below 220 K (Figs. 1 and 2).
However, the substructure is only slightly distinguished in the
low-frequency slopes. The frequencies of the $\delta$(OH) band
satellites at 99 K are 1148, 1120, 1095, 1068, 1040, 1011, 982,
956, and 932 cm$^{-1}$. The differences between the succeeding
peaks in this progression are almost constant that equals to $24.5
\pm 6$ cm$^{-1}$. The $\nu$(OH) band satellites in the spectra
below 220 K are located at the frequencies 2780, 2670, 2560, and
2445 cm$^{-1}$ and the respective difference between the maximums
is near $120 \pm 10$ cm$^{-1}$. When passing from higher to lower
temperatures, all the bands become sharper. At the same time their
relative positions and intensities are still retained. Note that a
similar band progression has also been observed for the $\nu$(OH)
and $\delta$(OH) bands in the spectra of $\alpha$-HIO$_{3}$ [15]
and $\alpha$-KIO$_{3}\cdot$HIO$_{3}$ [16] crystals.

The origin of such an unusual spectroscopic feature cannot be
associated with any possible strong anharmonic coupling between
the high-frequency vibrational modes of hydrogen bonds and
low-frequency lattice phonons as it was described in Ref. [26].
According to their theoretical model the appearance of a multiband
substructure at the low frequency slope of the $\nu$(OH)
absorption band in a solid is possible only in situations of
strong and very strong hydrogen bonds. However, this is not the
case for the crystal in question where only weak and medium strong
hydrogen bonds are present.

Taking into account the complex structure of the given crystal ($Z
= 24$), proton order for $T < T_{c} $ and the data on practically
independence of proton dynamics in solids, Fillaux et al. [17-22],
a radically new theory of the proton behavior can be proposed.

\section*{3. Clusterization in the framework of orthodox statistical
mechanics}

\hspace*{\parindent} Having considered whether nonhomogeneous
states, i.e. clusters, may appear spontaneously in systems with
hydrogen bonds, we first should describe the methodology developed
in Refs. [3-5] and then apply it to the case, which falls within
the Boltzmann statistics. We shall start from the construction of
the Hamiltonian for a system of interacting particles. The
Hamiltonian can be represented as follows
\begin{equation}
\label{2} H\left( {n} \right) = \sum\limits_{s} {E_{s}}  n_{s} -
\tfrac{{1}}{{2}}\sum\limits_{s,{\kern 1pt} {\kern 1pt} {s}'}
{V_{s{\kern 1pt} {s}'} {\kern 1pt} n_{s} {\kern 1pt} n_{{s}'} +
\tfrac{{1}}{{2}}\sum\limits_{s,{\kern 1pt} {\kern 1pt} {s}'}
{U_{s{\kern 1pt} {s}'} {\kern 1pt} n_{s} {\kern 1pt} n_{{s}'}} }
\end{equation}
where \textit{E}$_{s}$ is the additive part of the particle energy
in the \textit{s}th state. The main point of the approach is the
presentation of the total atom/molecular pair potential into two
terms: the repulsion and attraction components. So, in the
Hamiltonian (2) \textit{V}$_{ss'}$ and \textit{U}$_{ss''}$ are
respectively the paired energies of attraction and repulsion
between particles located in the states \textit{s} and $s^{\kern
1pt \prime}$. It should be noted that the signs before the
potentials in expression (2) specify proper signs of the
attractive and repulsive paired energies and this means that both
functions \textit{V}$_{ss'}$ and \textit{U}$_{ss'}$ in expression
(2) are positive. The statistical sum of the system
\begin{equation}
\label{3} Z = \sum\limits_{\{ n\}}  {\exp\left( { - H\left( {n}
\right)/k_{{\kern 1pt}\rm B} T} \right)}
\end{equation}
may be presented in the field form
\begin{eqnarray}
\label{4}  Z &=& \int_{ - \infty} ^{\infty}  D{\kern 1pt} \phi
\int_{ - \infty} ^{\infty} D{\kern 1pt} \psi \sum\limits_{\{ n\} }
\exp \Big[ \sum\limits_{s} \Big( \psi_{s} + i\phi_{s} \Big){\kern
1pt} {\kern 1pt} n_{s}
\\  \nonumber
&-&\tfrac 12 \sum\limits_{s,{\kern 1pt} {s}'} \Big(
{\tilde {U}_{s{\kern 1pt} {s}'}^{-1} {\kern 1pt} \phi_{s} {\kern
1pt} \phi _{{s}'} + \tilde {V}_{s{\kern 1pt} {s}'}^{-1} {\kern
1pt} \psi_{s} {\kern 1pt} \psi _{{s}'}} \Big) \Big]
\end{eqnarray}
due to the following representation known from the theory of Gauss
integrals
\begin{eqnarray}
\label{5} &&\exp\left( {\frac{{\rho
^{2}}}{{2}}\sum\limits_{s,{\kern 1pt} {\kern 1pt} {s}'}
{\mathcal{W}_{s{\kern 1pt} {s}'} {\kern 1pt} n_{s} {\kern 1pt}
n_{{s}'}} }  \right)      \\     \nonumber &&\quad\quad= {\rm Re}
\int_{-\infty}^{\infty} D{\kern 1pt} {\kern 1pt} \vartheta
\exp\left[ {\rho \sum\limits_{s} {n_{s} \vartheta _{s} -
\tfrac{{1}}{{2}}\sum\limits_{s,{\kern 1pt} {\kern 1pt} {s}'}
{\mathcal{W}_{s{\kern 1pt} {s}'}^{ - 1} \vartheta _{s} \vartheta
_{{s}'}} } } \right]
\end{eqnarray}
where $D{\kern 1pt} {\kern 1pt} \vartheta \equiv
\prod\nolimits_{s} {\sqrt {{\rm det}||\mathcal{W}_{s{\kern 1pt}
{s}'} ||}}  \sqrt {2\pi}  {\kern 1pt} d{\kern 1pt} \vartheta _{s}
$ implies the functional integration with respect to the field
$\vartheta $; $\rho ^{2} = \pm 1$ in relation to the sign of
interaction (+1 for attraction and $ - 1$ for repulsion). The
dimensionless energy parameters $\tilde {V}_{s{\kern 1pt} {s}'} =
V_{s{\kern 1pt} {s}'} /k_{\rm {\kern 1pt}B} T, \quad \tilde
{U}_{s{\kern 1pt} {s}'} = U_{s{\kern 1pt} {s}'} /k_{\rm {\kern
1pt}B} T,$ and $\tilde {E}_{s} /k_{\rm {\kern 1pt}B} T$ are
introduced into expression (4). Further, we will use the known
formula
\begin{equation}
\label{6} \frac{{1}}{{2\pi {\kern 1pt} i\quad} }\oint {d{\kern
1pt} z} {\kern 1pt} {\kern 1pt} z^{{\kern 1pt} {\kern 1pt} N - 1 -
\sum\nolimits_{s} {n_{s}} } = 1,
\end{equation}
which makes it possible to settle the quantity of particles in the
system, $\sum\nolimits_{s} {n_{s} = N} ,$ and, consequently, we
can pass to the consideration of the canonical ensemble of
\textit{N} particles. Thus the statistical sum (4) is replaced for
\begin{equation}
\label{7}
\begin{array}{l}
Z = {\rm Re}\frac 1{2\pi i}\oint {d{\kern 1pt} z} \int {D\phi \int
{D\psi {\kern 1pt} \exp \{ - \tfrac{{1}}{{2}}\sum\limits_{s,{\kern
1pt} {\kern 1pt} {s}'} {\left( {\tilde {U}_{s{\kern 1pt} {s}'}^{ -
1} {\kern 1pt} {\kern 1pt} \phi _{s} {\kern 1pt} \phi _{{s}'} +
\tilde {V}_{s{\kern 1pt} {s}'}^{ - 1} {\kern 1pt} \psi _{s} {\kern
1pt} \psi _{{s}'}} \right)}} }                      \\
 \quad \quad + \left( {N - 1} \right){\kern 1pt} {\kern 1pt}
{\kern 1pt} {\kern 1pt} \ln z \} \times \sum\limits_{\{ n_{s} \} =
0}^{1} { \exp \{ \sum\limits_{s} {n_{s} \left( {\psi _{s} +
i{\kern 1pt} \phi _{s} - \tilde {E}_{s}}  \right)} - \ln z\}}  .
\\
 \end{array}
\end{equation}

Summing over \textit{n}$_{s}$ we obtain
\begin{equation}
\label{8} Z = {\rm Re} \frac{{1}}{{2\pi i}}\int {D\phi \int {D\psi
\oint {d{\kern 1pt} z}} } {\kern 1pt} {\kern 1pt} e^{{\kern 1pt}
{\kern 1pt} S\left( {\phi ,{\kern 1pt} {\kern 1pt} \psi ,{\kern
1pt} {\kern 1pt} z} \right)}
\end{equation}
where
\begin{equation}
\label{9}
\begin{array}{l}
 S = \sum\limits_{s} {\left\{ { - \tfrac{{1}}{{2}}\sum\limits_{{s}'} {\left(
{\tilde {U}_{s{\kern 1pt} {s}'}^{ - 1} {\kern 1pt} {\kern 1pt} \phi _{s}
{\kern 1pt} \phi _{{s}'} + \tilde {V}_{s{\kern 1pt} {s}'}^{ - 1} {\kern 1pt}
\psi _{s} {\kern 1pt} \psi _{{s}'}}  \right)}}  \right.} \\
 \left. {\quad \quad + \eta {\kern 1pt} {\kern 1pt}  \ln \left| {1
+ \frac{{\eta} }{{z}}e^{{\kern 1pt} - \tilde {E}_{s}} e^{{\kern
1pt} \psi _{s}} \cos\phi _{s}}  \right|} \right\} + \left( {N - 1}
\right){\kern 1pt} {\kern 1pt} {\kern 1pt} \ln z. \\
 \end{array}
\end{equation}
Here, the symbol ç characterizes the kind of statistics: Bose
($\eta = +1$) or Fermi ($\eta = - 1$). Let us set $z = \xi +
i{\kern 1pt}\zeta $ and consider the action $S$ on a transit path
that passes through the saddle-point at a fixed imaginable
variable ${\rm Im} z = \zeta _{{\kern 1pt} 0} .$ In this case $S$
may be regarded as the functional that depends on the two field
variables $\phi $ and $\psi$, and the fugacity $\xi = e^{ - \mu
/k_{{\kern 1pt}\rm B} T}$ where $\mu $ is the chemical potential.

In a classical system the mean filling number of the \textit{s}th energy
level obeys the inequality
\begin{equation}
\label{10} n_{s} = \frac{{1}}{{\xi} }e^{ - \tilde {E}_{s}}  =
e^{\left( {\mu - E_{s}} \right)/k_{\rm {\kern 1pt}B} T} < < 1
\end{equation}
(note the chemical potential $\mu < 0$ and $|\mu |/k_{\rm {\kern
1pt}B} T \gg 1$). By this means, we can simplify expression (9)
expanding the logarithm into a Taylor series in respect to the
small second member. As a result, we get the action that describes
the ensemble of interacting particles, which are subjected to the
Boltzmann statistics
\begin{equation}
\label{11}
\begin{array}{l}
 S \cong \sum\limits_{s} {\left\{ { - \tfrac{{1}}{{2}}\sum\limits_{{s}'}
{\left( {\tilde {U}_{s{\kern 1pt} {s}'}^{-1} {\kern 1pt} {\kern
1pt} \phi _{s} {\kern 1pt} \phi _{{s}'} + \tilde {V}_{s{\kern 1pt}
{s}'}^{ - 1} {\kern 1pt} \psi _{s} {\kern 1pt} \psi _{{s}'}}
\right) + \frac{{1}}{{\xi} }e^{ - \tilde {E}_{s}} e^{\psi _{s}}
\cos \phi _{s}} }  \right\}}      \\    \quad \quad + \left( {N -
1} \right) \ln \xi . \\
 \end{array}
\end{equation}

The extremum of functional (11) is realized at the solutions of
the equations $\delta {\kern 1pt} S/\delta \phi = 0, \quad \delta
{\kern 1pt} S/\delta {\kern 1pt} \psi = 0,$ and $\delta {\kern
1pt} S/\delta \xi = 0$, or explicitly
\begin{equation}
\label{12}  \sum\limits_{{s}'} {\tilde {U}_{{s}'}^{ - 1}}  \phi
_{{s}'} = - \frac {1}{{\xi} }{\kern 2pt}e^{ - \tilde {E}_{s}}
e^{\psi _{s}} \sin\phi _{s} ,
\end{equation}
\begin{equation}
\label{13}  \sum\limits_{{s}'} {\tilde {V}_{{s}'}^{ - 1}}  \psi
_{{s}'} = - \frac {1}{{\xi} }{\kern 2pt} e^{ - \tilde {E}_{s}}
e^{\psi _{s}} \cos\phi _{s} ,
\end{equation}
\begin{equation}
\label{14}  \frac{{1}}{{\xi} }\sum\limits_{{s}'} {e^{ - \tilde
{E}_{{s}'}} e{\kern 1pt} ^{\psi _{{s}'}} \cos\phi _{{s}'}}  = N -
1.
\end{equation}

If we introduce the denotation
\begin{equation}
\label{15}  \mathcal{N}_{s} = \frac 1{\xi} {\kern 1pt} {\kern 1pt}
e^{ - \tilde {E}_{s}} e^{\psi _{s}} \cos\phi _{s},
\end{equation}
we will easily see from Eq. (15) that the sum $\sum\nolimits_{s}
{\mathcal{N}_{s}}  $ is equal to the number of particles in the
system studied. So the combined variable $\mathcal{N}_{s} $
specifies the quantity of particles in the \textit{s}th state.
This means that one may treat $\mathcal{N}_{s} $ as the variable
of particle number in a cluster. Using this variable, we can
represent the action (11) as a function of only one variable
$\mathcal{N}_{s}$ and the fugacity $\xi$
\begin{equation}
\label{16}
\begin{array}{l}
 S = - \sum\limits_{s,{\kern 1pt} {\kern 1pt} {s}'} {\left[ {\tilde
{V}_{s{\kern 1pt} {s}'} {\kern 1pt} \mathcal{N}_{s} {\kern 1pt}
\mathcal{N}_{{s}'} + \tilde {U}_{s{\kern 1pt} {s}'} {\kern 1pt}
\mathcal{N}_{s} {\kern 1pt} \mathcal{N}_{{s}'} \left( {\frac{{e^{
- 2\tilde {E}_{s} + 2\sum\nolimits_{{s}'} {\tilde {V}_{s{\kern
1pt} {s}'} \mathcal{N}_{{s}'}} } }}{{\xi ^{2}\mathcal{N}_{s}^{2}}
} - 1} \right)} \right]} \\
  \quad \quad + \sum\limits_{s} {\mathcal{N}_{s} {\kern
1pt} {\kern 1pt} \left( {1 + \ln\xi}  \right)}. \\
 \end{array}
\end{equation}

If we put the variable $\mathcal{N}_{s} = \mathcal{N} = \rm const$
in each of clusters, we may write instead of Eq. (14)
\begin{equation}
\label{17} \mathcal{N}K = N - 1.
\end{equation}
Here \textit{K} is the quantity of clusters and $\mathcal{N}$ is
classified as the number of particles in a cluster. Thus the model
deals with particles entirely distributed by clusters.

It is convenient now to pass to the consideration of one cluster
and change the discrete approximation to a continuous one. The
transformation means the passage from the summation over discrete
functions in expression (15) to the integration of continual
functions by the rule
\begin{equation}
\label{18}
\begin{array}{l}
 \sum\nolimits_{s} {f_{s} = \frac{{1}}{{\mathcal{V}}}{\kern 1pt}
\int\limits_{\rm cluster} {d{\kern 1pt} \vec {x}} {\kern 1pt}
f\left( {\vec {x}} \right){\kern 1pt} {\kern 1pt}}  =
\frac{{1}}{{\mathcal{V}}}{\kern 1pt} {\kern 1pt}
\int\limits_{0}^{2\pi} {d{\kern 1pt} {\kern 1pt} \phi}
\int\limits_{0}^{\pi } {d{\kern 1pt} \theta}  \sin \theta
\int\limits_{1}^{R/g} {d{\kern 1pt} {\kern 1pt} x} {\kern 1pt}
{\kern 1pt} x^{2}f\left( {x} \right) \\
 \quad \quad \quad {\kern 1pt} = {\kern 1pt} {\kern 1pt} {\kern 1pt}
3 \int\limits_{1}^{\mathcal{N}^{1/3}} {d{\kern 1pt} {\kern 1pt} x}
{\kern 1pt} {\kern 1pt} x^{2}f\left( {x} \right); \quad \quad
\quad
 \end{array}
\end{equation}
here $\mathcal{V} = 4\pi {\kern 1pt} g^{3}/3$ is the effective
volume of one particle, $R/{\kern 1pt} g$ is the dimensionless
radius of a cluster where \textit{g }is the mean distance between
particles in a cluster, and \textit{x} is the dimensionless
variable ($x = \left[ {1,{\kern 1pt} \;R/g} \right]$); therefore,
the number of particles in the cluster is linked with $R$ and $g$
by the relation $\left( {R/g} \right)^{3} = \mathcal{N}$ (see Ref.
[4] for details). Below we assume that $\mathcal{N} \gg 1$.

If we introduce the designations ($x$ is the dimensionless
variable)
\begin{equation}
\label{19} a = 3 {\kern 1pt} \int_{1}^{\mathcal{N}^{1/3}} {d{\kern
1pt} x{\kern 1pt} {\kern 1pt}}  x^{2}{\kern 1pt} \tilde
{U}(x),\quad \;\;\;\quad \quad \quad b = 3 {\kern 1pt}
\int_{1}^{\mathcal{N}^{1/3}} {d{\kern 1pt} x{\kern 1pt} {\kern
1pt}}  x^{2}{\kern 1pt} \tilde {V}(x),
\end{equation}
we will arrive at the action in the simple form
\begin{equation}
\label{20} S =K\cdot \Big\{ \left( {a - b} \right)
\mathcal{N}^{\kern 1pt 2} - \frac{{1}}{{\xi ^{2}}}{\kern 1pt}
a{\kern 1pt} e^{ - 2\tilde {E} + 2b\mathcal{N}} \Big\} + (N-1)
\ln\xi .
\end{equation}
The exponent term in the right-hand side of expression (20) is the
smallest one (owing to the inequality $e^{ - 2\tilde {E}}/\xi
{\kern 1pt} ^{2}{\kern 1pt} \ll 1$) and we omit it hereinafter.
Thus we shall start from the action
\begin{equation}
\label{21} S \simeq K \cdot \Big\{ \left( {a - b} \right)
\mathcal{N}^{\kern 1pt 2} +
 \mathcal{N} \ln\xi \Big\}.
\end{equation}

The extremum (minimum) of the action (21) is reached at the
meaning of $\mathcal{N},$ which is found from the equation $\delta
S/\delta \mathcal{N} = 0$ and satisfies the inequality $\partial
^{{\kern 1pt}2}S/\delta \mathcal{N}^{{\kern 1pt}2} > 0.$ The value
of $\mathcal{N}_{\rm extr} $ obtained in such a way will
correspond to the number of particles that compose a cluster.

\section*{4. Collective dynamics of hydrogen atoms}

\hspace*{\parindent} In the $\delta$-KIO$_3 \cdot$HIO$_3$ crystal
hydrogen atoms are not free, they are bound with oxygen atoms.
That is why we should analyze, what kinds of clusters could
hydrogen atoms form? In the first approximation, in the crystal
lattice atoms/molecules are characterized by their harmonic
vibrations in the vicinity of equilibrium positions. The amplitude
of small displacement \textit{A} of an atom in the crystal lattice
lies in the range 0.01 to 0.03 nm. In the crystal state this
parameter, i.e. \textit{A}, surrogates the de Broglie wavelength
$\lambda$ of a free atom. Since \textit{A} takes the place of
$\lambda$, the expression for the amplitude of inerton cloud (1)
becomes $\Lambda = A{\kern 1pt}c/\upsilon$. That is, $\Lambda \sim
1\;{\kern 1pt} {\kern 1pt} \mu$m and therefore the inerton cloud
of a crystal's entity can span over $10^{{\kern 1pt}3}g$, where
\textit{g} is the lattice constant. The general form of $A$ in the
crystal with $N$ atoms is $A_l = l / g N$ where $l = 2, {\kern
3pt}3, {\kern 3pt}..., (N-1)$, that is $A_l$ is the wavelength of
the $l$th mode of the acoustic spectrum of the crystal. Note that
such a strong overlap of inerton clouds of entities, in
particular, protons, is permanently used in the theory of
ferroelectrics for a long time: researchers implicitly introduce
an order parameter that introduces long-range interaction in the
proton subsystem below the phase transition that is needed for the
description of the ferro- or antiferroelectric state, though the
de Broglie wavelength of protons has only an order of the lattice
constant $g$ (see, e.g. Refs. [27,28]).

The methodology, which is described, does not need any fitting
order parameter. Instead, based on the findings of Fillaux et al.
[17-22] (especially see review article [21]), we can consider the
subsystem of hydrogen atoms as quite independent from the
framework. This is quite possible to do if we use the two-level
model for the description of hydrogen atoms; in the model each
hydrogen atom is able to occupy either the ground state or only
one excited state. Being in the excited state, the hydrogen atoms
will behave as typical quasi-particles, which are able to interact
each other. Due to the link with oxygens, the interaction between
the hydrogen atoms is rather Van der Waals' and can be simulated
by the Lennard-Jones potential. The second type of interaction
does not fall within the electromagnetic interaction and can be
assigned to the pure quantum mechanical interaction caused by the
overlapping of the hydrogen atoms' matter waves, i.e. elastic
inerton clouds.

Thus, the pair potential between hydrogen atoms can be written as
the sum of the Lennard-Jones potential and an additional harmonic
potential that takes into account the weak quantum mechanical
interaction of hydrogen atoms via the inerton field on the scale
$r \ll \Lambda $:
\begin{equation}
\label{eq22} W = \epsilon \left[ {\left( {\frac{{1}}{{r/g}}}
\right)^{12} - {\kern 1pt} {\kern 1pt} {\kern 1pt} {\kern 1pt}
\left( {\frac{{1}}{{r/g}}} \right)^{{\kern 1pt} 6}} \right] +
\tfrac 12{\kern 1pt}{\kern 1pt}\gamma {\kern 1pt} r^{{\kern 1pt}
2}.
\end{equation}
Here $\epsilon $ is the energy constant, $g$ is the equilibrium
distance between a pair of interacting hydrogen atoms and $\gamma
$ is the elasticity constant that characterizes the elastic
interaction of hydrogen atoms' inerton clouds (and may be partly
the backbone atoms).

Once again, why the potential is chosen in the form $\tfrac
12{\kern 1pt}\gamma {\kern 1pt} r^{{\kern 1pt} 2}$? This is
potential energy that characterizes oscillations of H-atoms near
their equilibrium positions; here $r$ is the difference between
positions of a pair of H-atoms. Therefore in this expression $r$
plays a role of the de Broglie wavelength of a H-atom, or a group
of collective vibrating H-atoms (in the latter case $r$ represents
a typical acoustic wavelength).

Let us now separate repulsive and attractive parts from the
potential (22) and represent them in the dimensionless form
\begin{equation}
\label{23} \tilde {U}\left( {x} \right) = \frac{{\epsilon}
}{{k_{{\kern 1pt}\rm B} T}}{\kern 1pt }\frac{{1}}{{x^{12}}},
\end{equation}
\begin{equation}
\label{24} \tilde {V}\left( {x} \right) = \frac{{\epsilon}
}{{k_{{\kern 1pt}\rm B} T}}{\kern 1pt }\frac{{1}}{{x^6}} -
\frac{{\gamma {\kern 1.5pt} \delta g^2}}{{2{\kern 1pt}k_{{\kern
1pt}\rm B} T}}{\kern 1.5pt} x^{2}
\end{equation}
where $\delta g$ is the amplitude of local deviation of a H-atom
from its equilibrium position (once again, $\delta g$ is the de
Broglie wavelength of an atom in a solid).

Substituting potentials (23) and (24) into the corresponding
formulas for parameters \textit{a} and \textit{b} (19), we get
instead of expression (21) with accuracy to $1/\mathcal{N}$
\begin{equation}
\label{25} S = K \cdot \Big\{ \Big( 36 \frac {\epsilon}{k_{\rm B}
T} - \frac {3}{10} \frac {\gamma {\kern 1.5pt} \delta g^{{\kern
1pt} 2}}{k_{\rm B} T} \mathcal{N}^{{\kern 1pt} 5/3} \Big)
\mathcal{N}^{{\kern 1pt} 2} + \mathcal{N} \ln\xi \Big\}.
\end{equation}

If we retain the major terms in the equation $\delta S/\delta
\mathcal{N} = 0,$ we will obtain retaining highest terms
\begin{equation}
 \label{26}
72{\kern 1pt} {\kern 1pt} \frac {\epsilon } {k_{\rm B} T} {\kern
1pt} \mathcal{N} - \frac{11}{10} {\kern 1pt} \frac {\gamma {\kern
1.5pt} \delta g^{2}} {k_{\rm B} T} {\kern 1pt} \mathcal{N}^{{\kern
2pt} 8/3} \cong 0.
\end{equation}

The solution to equation (26) is
\begin{equation}
\label{27} \mathcal{N} = \left( \frac {11}{720}{\kern 2pt} \frac
{\epsilon} {\gamma {\kern 1.5pt} \delta g^{2}} \right)^{3/5}.
\end{equation}
Here the interaction energy between a pair of hydrogen atoms is
taken to be typical Van der Waals', i.e. $\epsilon \simeq 10$
kJ/mol $ \equiv 1.66 \times 10^{ - 20}$ J; the amplitude of
deviation of a H-atom from the equilibrium position in a hydrogen
cluster studied can be put equal to $\delta g = 0.2 \times
10^{-10} $ m; the elasticity constant of the hydrogen atoms'
inerton field, i.e. $\gamma$, is a free parameter; let us set
$\gamma = 8.8$ N/m (on average, the force constant in a solid
comes out to several tens of N/m).

Substituting the aforementioned numerical values of $\epsilon$,
$\delta g$ and $\gamma$ into expression (27) we obtain:
$\mathcal{N} = 52$.

The value of hydrogen atoms $\cal N$ is also evident from the
spectra shown in Figs. 1 and 2. Indeed, it is obvious that our
cluster is some kind of a dynamic system, because hydrogen atoms,
which enter the cluster, do not move from their equilibrium
positions, but only trade inerton excitations. In other words, we
may treat the cluster as a crystallite in which H-atoms vibrate by
the rule typical for an usual crystallite: The kinetic energy of
H-atoms periodically passes to the potential one. On the
microscopic level this means that the inert mass of each of the
hydrogen atoms fluctuate periodically changing from $m_{{\kern
1pt} {\rm H}}-\frac 12 m^*$ to $m_{{\kern 1pt} {\rm H}} + \frac 12
m^{\ast}$, where $m^{\ast }$ is the effective mass of an inerton
excitation in the crystallite studied (in other words, the
vibratory motion of hydrogen atoms might be modulated by the
oscillation of the mass of the crystallite lattice's sites).

Inertons carry mass and feature velocity. Because of that, the
behavior of inerton excitations in the crystallite can be studied
starting from the Lagrangian that is typical for the crystal
lattice of a solid (see, e.g. Ref. [30])
\begin{equation}
\label{28} L = \frac{m_{\rm H}}{2}\sum\limits_{{\bf l};{\kern 1pt}
{\kern 1pt} {\kern 1pt} \alpha}  {\dot {u}_{{\kern 1pt} {\bf
l}{\kern 1pt} \alpha} ^{2} -}  {\kern 1pt} {\kern 1pt} {\kern 1pt}
\frac 12{\kern 1pt} {\kern 1pt} {\sum\limits_{{\bf l},{\bf
n};{\kern 1pt} \alpha ,\beta}}^{\prime} {\kern 2pt} {\kern 1pt}
\gamma _{\alpha {\kern 1pt} \beta}  {\kern 1pt} \left( {{\bf l} -
{\bf n}} \right){\kern 1pt} {\kern 1pt} u_{{\kern 1pt} {\bf
l}{\kern 1pt} \alpha}  u_{{\kern 1pt} {\bf n}{\kern 1pt} \beta}.
\end{equation}
Here the first term describes the kinetic energy of H-atoms and
the second term specifies their potential energy. The prime at the
sum symbol in Lagrangian (28) signifies that terms with coinciding
indices \textbf{l} and \textbf{n} are not taken into account in
summation. $u_{\kern 1pt \bf l {\kern 1pt \kern 1pt} \alpha}$ $(
\alpha = 1, {\kern 1pt} 2, {\kern 1pt} 3 )$ are three components
of displacement of a H-atom from the crystallite site whose
equilibrium position is determined by the lattice vector
\textbf{l}; $\dot {u}_{{\kern 1pt} {\bf l}{\kern 1pt \kern 1pt}
\alpha}$ are three components of the velocity of the H-atom;
$\gamma _{\alpha {\kern 1pt} \beta} {\kern 1pt} \left( {\bf l - n}
\right)$ are the components of the elasticity tensor of the
crystallite lattice and then the elasticity constant $\gamma$ that
enters into expressions from (22) to (27) can be considered as the
convolution of the tensor $\gamma_{\alpha \beta}$.

Let us now approximate the elasticity tensor $\gamma_{\alpha
\beta}$ by two components: longitudinal, $\gamma_{||}$, and
transversal, $\gamma_ {\bot}$.

The Lagrangian (28) produces the dispersion equation for allowed
frequencies of excitations in a cluster. The form of the equation
is the same as that for pure acoustic phonon branches in a solid
(see e.g. Ref. [29])
\begin{equation}
\label{29} \omega_{\ell,\kern 1pt j} \cong q_{j} {\kern 2pt} g
{\kern 1pt} \sqrt {\gamma_{\ell}{\kern 0.5pt} / m_{\rm H}} = q_{j}
{\kern 1pt} {\kern 2pt} v_{\ell}
\end{equation}
where $g \approx 0.3$ nm is the lattice constant (the typical
distance between hydrogen atoms in a hydrogen cluster); the index
$\ell$ characterizes the two types of inerton modes: longitudinal,
$\ell = ||$, and transversal, $\ell = \bot$. Hence $v_{||{\kern
1pt}( \bot )}$ the typical longitudinal and transversal sound
velocities in the crystal. Eq. (29) is correct until the
inequality $q_{j}{\kern 2pt } g < \pi $ holds where wavenumbers
$q_{\kern 0.5pt j} = \pi j/g\mathcal{N}$ and $j = \pm 1,{\kern
1pt} \pm 2,{\kern 1pt} {\kern 1pt} {\kern 1pt} ...,{\kern 1pt}
{\kern 1pt} {\kern 1pt} \pm \left( {\mathcal{N} - 1} \right)$.

The dispersion law (29) is still proper in the center of the
Brillouin zone. Figs. 1 and 2 depict the infrared spectra just in
the given range. A series of submaximums, which are superimposed
on the carrying absorption band, are equidistant. It is this
behavior that the spectrum of inerton excitations (29) prescribes.
Besides, as it follows from the form of the Lagrangian (28), the
equation of motion of each H-atom is reduced to the equation of
motion of a linear oscillator. Therefore, the coefficient of
absorption of an excitation is the same as that of absorption of
an oscillator. Thus, the rough estimation of the total absorption
coefficient of the vibratory hydrogen atoms can be represented as
follows
\begin{equation}
\label{30} K_{ \ell }\left( {\omega}  \right) \propto C_1 \left[
{\left( \omega - \omega _{\rm max, \kern 1pt \ell } \right)^{2} +
 \eta _{\ell {\kern 1pt} 1}^{\kern 1pt 2}} \right]^{{\kern
1pt} - 1} + C_2 \sum_{j=1}^{\cal N} \left[ {\left( {\omega -
\omega_{{\kern 0.5pt} \ell \kern 0.8pt {\kern 1pt} j}} \right)^{2}
+ \eta _{\ell {\kern 2pt} 2}^{\kern 1pt 2}} \right]^{{\kern 1pt} -
1}
\end{equation}

\noindent where $\omega _{{\rm max} {\kern 1pt} ||}$ corresponds
to the maximum of the spectra of $\nu$(OH) band and $\omega _{{\rm
max} {\kern 1pt} \bot }$ corresponds to the maximum of
$\delta$(OH) band shown in Figs. 1 and 2, respectively. The second
summand in the left-hand side of expression (30) describes the
absorption caused by inerton excitations, hence $\omega_{\ell
\kern 0.5pt j}$ obeys dispersion equation (29). Coefficients
$C_{1(2)}$ are constants.

If we build up the profiles of the absorption spectra by
expression (30), we will obtain graphs (Figs. 3 and 4),  which
agree closely with those shown in Figs. 1 and 2.
\begin{figure}
\begin{center}
\includegraphics[scale=0.33]{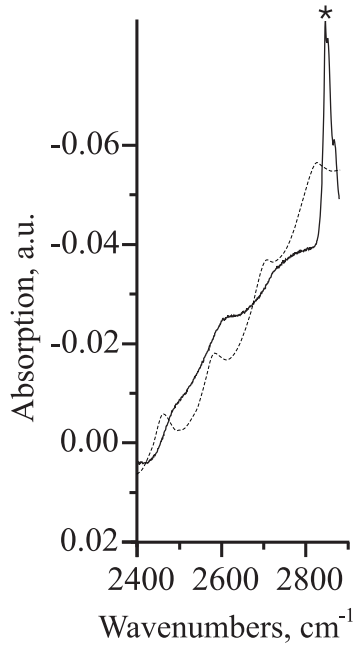}
\caption{Experimental curve (curve 3 of Fig. 1, the asteric
denotes the Nujol absorption band) and the curve calculated by
expression (30), or more exactly by formula
  $
K_{||}=C \times [(\omega - 2900)^2 + 300^2]^{-1} + \frac 1{800}
\sum_{j=0}^{51} [(\omega -120 \times (0.5 + j))^2 + (12+0.8
j)^2]^{-1}
  $ (dashed line).} \label{Figure 3}
\end{center}
\end{figure}

\begin{figure}
\begin{center}
\includegraphics[scale=0.4]{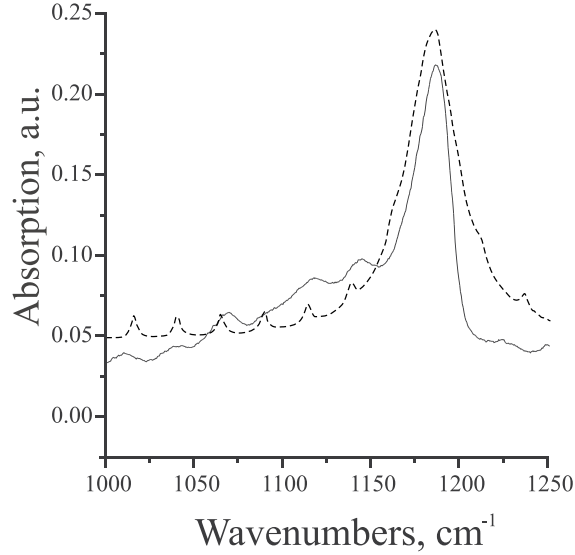}
\caption{Experimental curve (curve 5 of Fig. 2, solid line) and
the curve calculated by expression (30), or more exactly by
formula
  $
K_{\perp}= [(\omega - 1185)^2 + 20^2]^{-1} + \frac 1{800}
\sum_{j=0}^{51} [(\omega -24.5 \times (0.5 + j))^2 + (0.1 + 0.08
j)^2]^{-1}
  $ (dashed line).} \label{Figure 4}
\end{center}
\end{figure}

We can check on solution (27) because the same value of
$\mathcal{N}$ should follow from the comparison of the difference
between absorption frequencies $\Delta {\kern 1pt} \omega_{ \ell }
$ of two nearest submaximums written, on the one hand, for
incident photons and, on the other hand, for excited inerton
excitations. In fact, as it is evident from Figs. 1 and 2,
\begin{equation}
\label{eq31} \Delta {\kern 1pt} \omega_{||} {\kern 1pt} =
120\,\,{\kern 1pt} {\kern 1pt} {\rm cm}^{ - 1} \times {\kern 1pt}
{\kern 1pt} {\kern 1pt} c{\kern 1pt} ;\quad \quad \quad {\kern
1pt} \,\,\,\Delta {\kern 1pt} \omega_{ \bot } = 24.5\,\,\,{\rm
cm}^{ - 1} \times {\kern 1pt} {\kern 1pt} {\kern 1pt} c
\end{equation}

\noindent where \textit{c} is the velocity of light. At the same
time, it is evident from the dispersion law (31) that these values
also come out to
\begin{equation}
\label{32} \Delta {\kern 1pt} \omega_{\ell} =  v_{ \ell }{\kern
2pt} \Delta q
\end{equation}
where allowance is made for the explicit form of the wavenumber
$q_{j}$ in the cluster (i.e. $\Delta q = 2{\kern 1pt}\pi/ g{\cal
N}$). Substituting values of $\Delta {\kern 1pt} \omega_{\ell}$
from expressions (31) into Eq. (32), choosing the corresponding
fitting values of the sound velocities $v_{{\kern 0.5pt} ||} = 4.5
\times 10^{{\kern 1pt} 3}$ m/s and $v_{{\kern 0.5pt} \bot } = 1.8
\times 10^{3}$ m/s, and setting the crystallite constant $g = 3
\times 10^{ - 10}$ m, we will arrive at the same value of hydrogen
atoms in a cluster that expression (27) prescribes, $\mathcal{N} =
52$.

\section*{5. Concluding Remarks}

\hspace*{\parindent} Thus, the IR spectroscopic study of the
$\delta$-KIO$_3 \cdot$HIO$_3$ crystal [1] exhibited the band fine
structures in the stretching and bending vibrations of hydrogen
bonds. The appearance of those substructures has been associated
with excited states of hydrogen atoms in clusters, which are
formed due to the mechanism stated above. Hydrogen atoms, which
enter a cluster, do not move from their equilibrium positions in
the crystal backbone, however, the interaction of the hydrogen
atoms through the electromagnetic field and the quantum mechanical
(i.e. inerton) field brings into existence their synchronous
vibrations. Thereby, a fundamentally new type of the interaction
between hydrogen atoms, which result in their collective
oscillations, has been revealed. Since inertons are carriers of
mass, we may say that the atomic mass of hydrogen atoms undergoes
periodical oscillations in clusters. The described experiment is
of infrequent occurrence, though it has already been known
previously: Podkletnov's phenomenon [31,32], i.e. loss of a part
of the mass of electrons in the superconductor state at special
conditions, serves as a good example of soundness of the theory
proposed.

The estimated number of hydrogen atoms ${\kern 1pt} \mathcal{N} =
52$ correlates well with the appropriate number calculated from
the consideration of the band fine structures.

   The phenomenon studied in the present work in fact has
demonstrated the inner nature of the interaction of simplest
entities, i.e. hydrogen atoms, in the solid. Since the inerton
field concerns the matter construction in general, the results
obtained in this work might be very required at the investigation
of other difficult problems associated with a subtle analysis of
physical processes in condensed media.

\section*{Acknowledgment}

\hspace*{\parindent} I am indebted to Professor F. Fillaux and
Professor G. Zundel for the useful discussion of the results
presented in this work.


\medskip

\begin{thebibliography}{99}

\bibitem{1} B. Engelen, T. Gavrilko, M. Panthofer, G. A. Puchkovska, J. Baran,
             and H. Ratajczak, \textit{ Selected papers from the
             Int. Conf. on Spectroscopy of Molecules
             and Crystals}, ed.: G.A. Puchkovska.
             {\it Proceed. of SPIE}, {\bf 4938}, 15-20 (2002).

\bibitem{2} E. D. Belotsky, and B. I. Lev, \textit{Theor. Math. Phys}.
         \textbf{60}, 120 (1984); in Russian.

\bibitem{3} V. Krasnoholovets, and B. Lev., \textit{Ukr. Fiz. Zh.}
         \textbf{33}, 296 (1994); in Ukrainian.

\bibitem{4} B. I. Lev, and A. Yu. Zhugaevich, \textit{Phys. Rev. E}
          \textbf{57}, 6460 (1998).

\bibitem{5} V. Krasnoholovets, and B. Lev, \textit{Cond. Matt. Phys}.
          {\bf 6}, 67 (2003) (also arXiv.org e-print archive
           cond-mat/0210131).

             under consideration (also arXiv.org e-print archive,
             cond-mat/0210131).

\bibitem{6} V. Krasnoholovets, and D. Ivanovsky, \textit{Phys. Essays}
             \textbf{6}, 554 (1993) (also arXiv.org e-print
             archive, quant-ph/9910023).

\bibitem{7} V. Krasnoholovets, \textit{Phys. Essays} \textbf{10}, 407
             (1997) (also arXiv.org e-print archive quant-ph/9903077).

\bibitem{8} V. Krasnoholovets, and V. Byckov, \textit{Ind. J. Theor.
             Phys}. \textbf{48}, 1 (2000) (also quant-ph/0007027).

\bibitem{9} V. Krasnoholovets, \textit{Ind. J. Theor. Phys}. \textbf{48},
             97 (2000) (also quant-ph/0103110).

\bibitem{10} V. Krasnoholovets, \textit{Ind. J. Theor. Phys}. \textbf{49},
             1 (2001) (also quant-ph/9906091).

\bibitem{11} V. Krasnoholovets, \textit{Ind. J. Theor. Phys}.\textbf{ 49},
            85 (2001) (also quant-ph/9908042).

\bibitem{12} V. Krasnoholovets, {\it Spacetime \& Substance}, \textbf{1}, 172
             (2000) (also quant-ph/0106106); \textit{Int. J. Comput. Anticipat.
              Systems}, {\bf 11}, 164 (2002) (also quant-ph/0109012).

\bibitem{15} V. Krasnoholovets, {\it Ann. de la Fond. L. de
              Broglie}, \textbf{27}, no. 1, 93 (2002)
              (also quant-ph/0202170).

\bibitem{14} B. Engelen, T. Gavrilko, M. Panh\"ofer, G. Puchkovska, and I.
             Sekirin, \textit{J. Mol. Struct}. \textbf{523}, 163 (2000).

\bibitem{15} T. Gavrilko, G. Puchkovska, Yu. Polivanov, and A. Yaremko,
             \textit{Ukr. Fiz. Zh}. \textbf{30}, 29 (1985); in Russian.

\bibitem{16} A. Barabash, J. Baran, T. Gavrilko, K. Eshimov, G. Puchkovska, and
             H. Ratajczak, \textit{J. Mol. Struct}. \textbf{404}, 187 (1997).


\bibitem{17} F. Fallaux, J. Tomkinson, and J. Penfold, \textit{Cem. Phys}.
             \textbf{124}, 425 (1988).

\bibitem{18} F. Fallaux, J. P. Fontaine, M. H. Baron. G. J. Kearly, and J.
             Tomkinson, \textit{Chem. Phys}. \textbf{176}, 249 (1993).

\bibitem{19} F. Fallaux, J. P. Fontaine, M. H. Baron. G. J. Kearly, and J.
             Tomkinson, \textit{Biophys}. \textit{Chem}. \textbf{53}, 155
             (1994).

\bibitem{20} F. Fillaux, N. Leygue, J. Tomkinson, A. Cousson and W. Paulus,
             \textit{Chem. Phys}. \textbf{244}, 387 (1999).

\bibitem{21} F. Fillaux, \textit{J. Mol. Struct}. \textbf{511-512}, 35
             (1999).

\bibitem{22} F. Fillaux, \textit{Solid State Ionics} \textbf{125}, 69
             (1999).

\bibitem{23} T. Horsewill, M. Johnson, and H. P. Trommsdorff, Europhys.
             News \textbf{28}, 140 (1997).

\bibitem{24} Ch. Rambaud, and H. P. Trommsdorff, \textit{Chem. Phys. Lett}.
             \textbf{306}, 124 (1999).

\bibitem{25} S. Bratos, and H. Ratajczak, \textit{J. Chem. Phys}.
              \textbf{76}, 77 (1982).

\bibitem{26} H. Ratajczak, and A. M. Yaremko, \textit{Chem. Phys. Lett}.
             \textbf{243}, 348 (1995).

\bibitem{27} V. G. Vaks, \textit{Introduction to the microscopic theory of
             ferroelectrics} (Nauka, Moscow, 1973); in Russian.

\bibitem{28} R. Blinc, and B. \v{Z}ec\v{s}, \textit{Soft mode in
             ferroelectrics and antiferroelectrics} (North-Holland
             Publishing Company - Amsterdam, Oxford; American Elsevier
             Publishing Company, Inc. - New York, 1974).

\bibitem{29} V. Krasnoholovets, hep-th/0205196.

\bibitem{30} A. S. Davydov, \textit{The theory of solid} (Nauka, Moscow,
             1976); in Russian.

\bibitem{31} E. Podkletnov, cond-mat/9701074.

\bibitem{32} E. Podkletnov and G. Modanese, physics/0108005.

\end{thebibliography}
\end{document}